\documentclass{article}[11pt]
\usepackage{amsfonts}
\usepackage{CJK,amsmath,indentfirst}
\begin{document}

\title{\bf Interactions among different types of nonlinear waves described by the Kadomtsev-Petviashvili Equation $^{\dagger}$}

\author{\footnotesize Xue-Ping Cheng$^{1}$, Chun-Li Chen$^{2}$ and S. Y. Lou$^{1,3}$\thanks{Corresponding author.}\\
\footnotesize $^{1}$ \it Department of Physics, Shanghai Jiao Tong University, Shanghai, 200240, China\\
\footnotesize $^{2}$ \it Department of Mathematics, Shanghai Jiao Tong University, Shanghai, 200240, China\\
\footnotesize $^{3}$ \it Department of Physics, Ningbo University, Ningbo, 315211, China}
\date{}
\maketitle
\parindent=0pt
\textbf{Abstract:} In nonlinear physics, the interactions among solitons are well studied thanks to the multiple soliton solutions can be obtained by various effective methods. However, it is very difficult to study interactions among different types of nonlinear waves such as the solitons (or solitary waves), the cnoidal periodic waves and Painlev\'e waves. In this paper,  the nonlocal symmetries related to the Darboux transformations (DT) of the Kadomtsev-Petviashvili (KP) equation is localized after imbedding the original system to an enlarged one.
Then the DT is used to find the corresponding group invariant solutions such that interaction solutions among different types of nonlinear waves can be found. It is shown that starting from a Boussinesq wave or a KdV-type wave, which are two basic reductions of the KP equation, the essential and unique role of the DT is to add an additional soliton.  

\vskip.4in
\renewcommand{\thesection}{\arabic{section}}
\parindent=20pt
\section{Introduction}
To study the nonlinear waves, there are typical important prototype models such as the KdV (Korteweg de-Vries), KP (Kadomtsev-Petviashvili), NLS (nonlinear Schr\"odinger) and SG (sine-Gordon) models. These systems appear in almost all the physics fields, especially, in fluid physics, plasma physics, optics, condense matter, quantum fields and astrophysics \cite{KdV}. Many kinds of nonlinear waves (nonlinear excitations) such as the solitons, conoidal periodic waves, Painlev\'e waves (solutions of the Painlev\'e transcendent I--VI) are found by using various effective methods. However, to study the interactions among these nonlinear waves is still a thorny issue though the interactions among solitons have been studied well.

There are some excellent approaches to find multiple soliton solutions of integrable system. For instance, the simplest method may be the Hirota's bilinear method \cite{Hirota}. Usually, whence a nonlinear system is changed to a bilinear form, its multiple soliton solutions can be directly written down and then the interactions among solitons can be studied in detail from the analytic expressions. The Darboux transformations (DT) and/or the B\"acklund transformations (BT) are two other effective methods\cite{DT}. Using these two methods, in principle, one can obtain a new solution from a known one. However, in practice, one can only find multiple soliton solutions starting from simple constant solutions. It is still very difficult to find new explicit solutions starting from nonconstant nonlinear waves such as the cnoidal waves and Painlev\'e waves via DTs and BTs. It is not very clear that what kind of solutions can be find via DT and/or BT if one takes a nonconstant and non-soliton solutions as a seed.

Since the Lie group theory was introduced by Sophus Lie to study
differential equations \cite{Lie}, the study of Lie group has
always been an important subject in mathematics and physics. Using both
classical and non-classical Lie group approaches \cite{Olver,Bluman},
one can reduce dimensions of partial differential equations (PDEs)
and proceed to construct the analytical solutions of these PDEs.

Traditionally, to study symmetry groups one studies their symmetry algebras first and then the related groups via Lie's first theorem because the algebra problems are linear while the group problems are nonlinear.
The recent studies show us that for many kinds of nonlinear partial differential equation systems, to firstly find symmetry groups of the nonlinear systems may be much more convenient than to firstly find related symmetry algebras \cite{Group}.

Actually, there exist some other types of examples in soliton theory in which the groups (finite transformations) such as the
DTs and BTs have been
successfully studied while the corresponding symmetry algebras are not very familiar. In Refs. ${\cite{Lou1,Lou2}}$, the authors have shown us that the symmetries related to DTs for some well known
equations, such as the KdV equation, the KP equation, the CDGKS
equation, etc. are nonlocal.
It is known that using Lie point symmetries, the dimensions of the related PDEs can be reduced.
An important problem is how to use the DT related symmetries to reduce the dimensions of the original PDEs?
 A simple direct way to solve this problem is to prolong the system such that the nonlocal symmetry of the
 original system becomes a local one of the prolonged system \cite{Nonlocal}.
 In this paper, we will show that whence the DT related nonlocal symmetries are localized, it will be quite easy to obtain new exact solutions from non-soliton type seed solutions via DT related symmetry reduction approaches.

In Sec. 2 of this paper, the first DT related nonlocal symmetry of the KP equation is localized. In Sec. 3, using the localized symmetry of the first DT, the Levi transformation (the second DT) is re-obtained via Lie's first theorem.
In Sec.4, the similarity reductions for the prolonged system are
considered according to the standard Lie point symmetry approach. Some corresponding explicit interaction solutions among solitons and Boussinesq waves and KdV type waves are given in Sec. 5.
The last section is
devoted to a short summary and discussion.

\section{Localization of the non-local symmetry related to the first DT of the KP equation}
We consider the KP equation \cite{KP} in the form
\begin{equation}\label{21}
(u_{t}-6uu_{x}+u_{xxx})_{x}+3u_{yy}=0,\ \ \ \ \  \ \ \  \ \ \ \ \ \
\ \
\end{equation}
where the subscripts denote derivatives. The KP equation was firstly derived to study the evolution of long ion-acoustic waves of small amplitude propagating in plasmas under the effect of long transverse perturbations \cite{KP}. The KP equation was widely accepted as a natural extension of
the classical KdV equation to two spatial dimensions, and was later derived as a model for surface and internal water waves \cite{AS}, and in nonlinear optics \cite{PSK} and almost in all other physical fields such as in shallow water waves, ion-acoustic waves in plasmas, ferromagnetics, Bose-Einstein
condensation and string theory. The KP equation is also used as a classical model for developing and testing of new mathematical techniques.

It is well known that the KP
equation possesses the Lax pair
\begin{equation}\label{22a}
\hspace{-3.8cm}\psi_{xx}-u\psi+\psi_{y}=0,
\end{equation}
\begin{equation}\label{22b}
\psi_{t}+4\psi_{xxx}-6u\psi_{x}-3\left(u_{x}-\int u_{y}dx\right)\psi=0.
\end{equation}
and the adjoint Lax pair
Eq.(\ref{21})
\begin{equation}\label{23a}
\hspace{-3.8cm}\phi_{xx}-u\phi-\phi_{y}=0,
\end{equation}
\begin{equation}\label{23b}
\phi_{t}+4\phi_{xxx}-6u\phi_{x}-3\left(u_{x}+\int u_{y}dx\right)\phi=0.
\end{equation}
That is to say the integrable conditions of
Eqs.(\ref{22a},\ref{22b}) and (\ref{23a},\ref{23b}),
$\psi_{yt}=\psi_{ty}$ and $\phi_{yt}=\phi_{ty}$, are just the KP
equation (\ref{21}).

A symmetry, $\sigma^u$ of the KP equation is defined as a solution of its linearized equation
\begin{equation}\label{sym}
\left(\sigma^u_{t}-6\sigma^uu_{x}-6u\sigma^u_{x}+\sigma^u_{xxx}\right)_{x}+3\sigma^u_{yy}=0
\end{equation}
that means \eqref{21} is form invariant under the transformation
\begin{equation}\label{infu}
u\rightarrow u+\epsilon \sigma^u
\end{equation}
with the infinitesimal parameter $\epsilon$.

In Ref. \cite{Lou1}, the non-local symmetry corresponding to the
DT of the KP equation {\cite{Ma}} has been
derived, which is in the form
\begin{equation}\label{27}
\sigma^u=(\phi\psi)_{x},
\end{equation}
where $\psi,\phi$ satisfy the Lax pairs (\ref{22a},\ref{22b}) and
(\ref{23a},\ref{23b}).

Because \eqref{27} is just the infinitesimal form of the DT, one can naturally believe that it can be localized to a Lie point symmetry such that we can use the Lie's first theorem to recover the original DT.
From (\ref{27}), it can be apparently seen
that the non-local symmetry contains the space derivative of
functions $\psi$ and $\phi$. Then, to localize the nonlocal symmetry \eqref{27}, we have to know what are the corresponding transformations for the
quantities $\psi,\ \phi$,
\begin{equation}\label{p1}
\psi_1\equiv \psi_x,
\end{equation}
and
\begin{equation}
\label{p2}
\phi_1\equiv \phi_x
\end{equation}
whence the field $u$ has the symmetry transformation $u\rightarrow u+\epsilon \sigma^u$. In other words, we have to solve the linearized equations of \eqref{22a} , \eqref{23a}, \eqref{p1} and \eqref{p2},
\begin{eqnarray}
&&\sigma^\psi_{xx}-\sigma^u\psi-u\sigma^\psi+\sigma^\psi_{y}=0,\nonumber\\
&&\sigma^\phi_{xx}-\sigma^u\phi-u\sigma^\phi-\sigma^\phi_{y}=0,\nonumber\\
&&\sigma^{\psi_1}= \sigma^\psi_x,\nonumber\\
&&\sigma^{\phi_1}= \sigma^\phi_x\label{simphi}
\end{eqnarray}
whence $\sigma^u$ is given by \eqref{27}.

It is not difficult to verified that the solution of \eqref{simphi} with \eqref{27} has the form
\begin{eqnarray}
&&\sigma^\psi=\frac12\psi p,\nonumber\\
&&\sigma^\phi=\frac12\phi p,\nonumber\\
&&\sigma^{\psi_1}=\frac12(\psi_1p+\psi^2\phi),\nonumber\\
&&\sigma^{\phi_1}=\frac12(\phi_1p+\phi^2\psi),\label{pppp}
\end{eqnarray}
where the new quantity $p$ is defined as
\begin{eqnarray}
p_x=\psi\phi.\label{p}
\end{eqnarray}
Two compatibility conditions of \eqref{p} are worth to mentioned here
\begin{eqnarray}
p_y&=&\psi\phi_1-\psi_1\phi,\label{py}\\
p_t&=&4\psi_1\phi_1+6u\psi\phi-4\phi\psi_{1x}-4\psi\phi_{1x},\label{pt}
\end{eqnarray}
that means the conditions $p_{xt}=p_{tx},\ p_{ty}=p_{yt}$ and $p_{xy}=p_{yx}$ are satisfied identically.

The appearance of the quantity $p$ in the symmetry solution \eqref{pppp}, we have to further solve the linearized equation of \eqref{p}
\begin{eqnarray}
\sigma^p_x=\sigma^\psi\phi+\psi\sigma^\phi\label{sp}
\end{eqnarray}
with the condition \eqref{pppp}.

It is interesting that the solution of \eqref{sp} with \eqref{pppp} has the simple form
\begin{eqnarray}
\sigma^p=\frac12p^2. \label{spdt}
\end{eqnarray}

The result \eqref{spdt} hints us that $p$ is a solution of the Schwarzian KP equation,
\begin{eqnarray}
\left[\frac{p_t}{p_x}+\left\{p;\ x\right\}+ \frac32 \frac{p_y^2}{p_x^2}\right]_x+3\left(\frac{p_y}{p_x}\right)_y =0, \label{skp}
\end{eqnarray}
where the Schwarzian derivative $\{p;\ x\}\equiv \frac{p_{xxx}}{p_x}-\frac32
\frac{p_{xx}^2}{p_x^2}$ and the quantities $\frac{p_t}{p_x}$ and $\frac{p_y}{p_x}$ are all invariant under the M\"obious transformation invariant with the infinitesimal transformation \eqref{spdt}. This fact implies a new method to find Schwarzian forms of nonlinear systems which may be more significant in the future studies in discrete systems where the traditional Pailev\'e analysis \cite{WTC} and the singularity manifold do not work.

The results \eqref{pppp} and \eqref{spdt} show us that the DT related nonlocal symmetry \eqref{27} in the original space $\{x,\ y,\ t,\ u\}$ has been successfully localized to a Lie point symmetry in the enlarged space $\{x,\ y,\ t,\ u,\ \psi,\ \phi,$ $ \psi_1,\ \phi_1,\ p\}$ with the vector form
\begin{eqnarray}
V=(\psi\phi_1+\phi\psi_1)\partial_u+\frac12\psi p\partial_\psi+\frac12\phi p\partial_\phi+\frac12(\psi_1p+\psi^2\phi)\partial_{\psi_1}+\frac12(\phi_1p+\phi^2\psi)\partial_{\phi_1}
+\frac12p^2\partial_p. \label{sVdt}
\end{eqnarray}
In other words the DT related symmetry is only a special Lie point symmetry \eqref{sVdt} of the equation system \{(\ref{21}), (\ref{22a}), (\ref{23a}),
(\ref{p1}), (\ref{p2}), (\ref{p})\}. The equations \eqref{22b}, \eqref{23b}, \eqref{py} and \eqref{pt} may should also be included in the enlarged system. However, we only treat them as the compatibility conditions of (\ref{21}, \ref{22a}), (\ref{21}, \ref{23a}) and so on.

\section{Second type DT of the KP equation}
In this section, we study the finite transformation form of \eqref{sVdt} by using Lie's first theorem. To find the nonlocal symmetry \eqref{27} from DT, we take the infinitesimal group parameter is just the spectral parameter \cite{Lou2,Lou1}. So it is natural that the finite transformation of \eqref{sVdt} is just the usual first type of DT if one take the spectral parameter as group parameter. Here we are more interest to ask if we take the group parameter is independent of the spectral parameter, what kind of finite transformation will be found? That means we should solve the following ``initial value problem"
\begin{eqnarray}
\frac{d u'(\epsilon)}{d\epsilon}&=&(\psi\phi_1+\phi\psi_1)\qquad u'(0)=u,\nonumber\\
\frac{d \psi'(\epsilon)}{d\epsilon}&=&\frac12\psi'(\epsilon) p'(\epsilon), \qquad \psi'(0)=\psi,\nonumber\\
\frac{d \phi'(\epsilon)}{d\epsilon}&=&\frac12\phi'(\epsilon) p'(\epsilon), \qquad \phi'(0)=\phi,\nonumber\\
\frac{d \psi'_1(\epsilon)}{d\epsilon}&=&\frac12(\psi'_1(\epsilon) p'(\epsilon)+\psi'(\epsilon)^2 \phi'(\epsilon),
 \qquad \psi'_1(0)=\psi_1,\nonumber\\
\frac{d \phi'_1(\epsilon)}{d\epsilon}&=&\frac12(\phi'_1(\epsilon) p'(\epsilon)+\phi'(\epsilon)^2 \psi'(\epsilon),
 \qquad \phi'_1(0)=\phi_1,\nonumber\\
\frac{d p'_1(\epsilon)}{d\epsilon}&=&\frac12p'(\epsilon)^2 ,
 \qquad p'_1(0)=p,\label{initial}
\end{eqnarray}
with the group parameter $\epsilon$ being independent of the spectral parameter.

By solving the initial value problem \eqref{initial}, it is not difficult to find the following theorem. \\
\bf Theorem 1. \rm
If $\{u,\ \psi,\ \phi,\ \psi_1,\ \phi_1,\ p\}$ is a solution of the extended system \eqref{21}-\eqref{23b}, \eqref{p1}, \eqref{p2} and \eqref{p}, then so is $\{u'(\epsilon),\ \psi'(\epsilon),\ \phi'(\epsilon),\ \psi'_1(\epsilon),\ \phi'_1(\epsilon),\ p'(\epsilon)\}$ with
\begin{eqnarray}
p'(\epsilon)&=&\frac{2p}{2-\epsilon p},\nonumber\\
\psi'(\epsilon)&=&\frac{2\psi}{2-\epsilon p},\nonumber\\
\phi'(\epsilon)&=&\frac{2\phi}{2-\epsilon p},\nonumber\\
\psi'_1(\epsilon)&=&\frac{2\psi_1}{2-\epsilon p}+\frac{2\epsilon\psi^2\phi}{(2-\epsilon p)^2},\nonumber\\
\phi'_1(\epsilon)&=&\frac{2\phi_1}{2-\epsilon p}+\frac{2\epsilon\phi^2\psi}{(2-\epsilon p)^2},\nonumber\\
u'(\epsilon)&=&u+\frac{2\epsilon(\psi\phi_1+\phi\psi_1)}{2-\epsilon p}+\frac{2\epsilon^2\phi^2\psi^2}{(2-\epsilon p)^2}\label{DT2}
\end{eqnarray}
for arbitrary group parameter $\epsilon$.

It should be emphasized that this finite transformation theorem is not the usual DT (the first type of DT). It is equivalent of the so-called Levi transformation, the second type of Darboux transformations \cite{DT2}.
In other words, at the algebra level, two types of DTs share a same form but with different group parameters, the first one is just the spectral parameter while the second one is an independent parameter.

From the first equation of \eqref{DT2}, we know that the second DT is nothing but the M\"obious transformation of the Schwarzian forms of the original nonlinear systems.

\section{New symmetry reductions of the KP equation}
Now we take into account the Lie symmetries of the closed enlarged system. It
is known from the standard Lie point symmetry group approach that, in order
to find the Lie point symmetry, we may assume the symmetries
have the vector form
\begin{subequations}\label{30}
\begin{equation}
V=X\frac{\partial}{\partial x}+Y\frac{\partial}{\partial
y}+T\frac{\partial}{\partial t}+U\frac{\partial}{\partial u}+\Psi\frac{\partial}{\partial \psi}+\Phi\frac{\partial}{\partial \phi}+\Psi_{1}\frac{\partial}{\partial\psi_{1}}+\Phi_{1}\frac{\partial}{\partial
\phi_{1}}+P\frac{\partial}{\partial p},
\end{equation}
\end{subequations}
where $X,Y,T,U,\Psi,\Phi,\Psi_{1},\Phi_{1},P$ are the functions with
respect to $\{x,y,t,u,\psi,\phi,\psi_{1},\phi_{1},p\}$, which means
that the closed system is invariant under the transformations
\begin{equation}
\{x,y,t,u,\psi,\phi,\psi_{1},\phi_{1},p\}\rightarrow\{x+\epsilon
X,y+\epsilon Y,t+\epsilon T,u+\epsilon
U,\psi+\epsilon\Psi,\phi+\epsilon\Phi,\psi_{1}+\epsilon\Psi_{1},
\phi_{1}+\epsilon\Phi_{1},p+\epsilon P\}
\end{equation}
with a small parameter $\epsilon$. Equivalently, due to the fact
that the system is not explicitly space-time dependent, the
symmetries in the vector form (\ref{30}) can be written as a
function form
\begin{subequations}\label{31}
\begin{equation}
\sigma_{u}=Xu_{x}+Yu_{y}+Tu_{t}-U,\ \ \ \ \ \ \
\end{equation}
\begin{equation}
\sigma_{\psi}=X\psi_{x}+Y\psi_{y}+T\psi_{t}-\Psi,\ \ \ \ \ \
\end{equation}
\begin{equation}
\sigma_{\phi}=X\phi_{x}+Y\phi_{y}+T\phi_{t}-\Phi,\ \ \ \ \ \ \
\end{equation}
\begin{equation}
\sigma_{\psi_{1}}=X\psi_{1x}+Y\psi_{1y}+T\psi_{1t}-\Psi_{1},
\end{equation}
\begin{equation}
\sigma_{\phi_{1}}=X\phi_{1x}+Y\phi_{1y}+T\phi_{1t}-\Phi_{1},\
\end{equation}
\begin{equation}
\sigma_{p}=Xp_{x}+Yp_{y}+Tp_{t}-P. \ \ \ \ \ \ \ \
\end{equation}
\end{subequations}
In this notation,
$\sigma_{u},\sigma_{\psi},\sigma_{\phi},\sigma_{\psi_{1}},
\sigma_{\phi_{1}},\sigma_{p}$ are the solutions of the symmetry
equations, i.e., the linearized equations for the closed system
\begin{subequations}\label{32}
\begin{equation}
\sigma_{u,xt}-12u_{x}\sigma_{u}-6u\sigma_{u,xx}-6\sigma_{u}u_{xx}
+\sigma_{u,xxxx}+3\sigma_{u,yy}=0,
\end{equation}
\begin{equation}
\sigma_{\psi,xx} -u\sigma_{\psi}-\sigma_{u}\psi+\sigma_{\psi,y}=0,
\end{equation}
\begin{equation}
\sigma_{\phi,xx} -u\sigma_{\phi}-\sigma_{u}\phi-\sigma_{\phi,y}=0,\
\end{equation}
\begin{equation}
\hspace{-2.2cm}\sigma_{\psi_{1}}-\sigma_{\psi,x}=0,\
\end{equation}
\begin{equation}
\hspace{-2.3cm}\sigma_{\phi_{1}}-\sigma_{\phi,x}=0,\
\end{equation}
\begin{equation}
\hspace{-1.4cm}\sigma_{p}-\psi\sigma_{\phi}-\sigma_{\psi}\phi=0.\
\end{equation}
\end{subequations}
Substituting Eq.(\ref{31}) into Eq.(\ref{32}) and eliminating
$u_{yy}, \psi_{xx},\phi_{xx},\psi_{x},\phi_{x}$ and $p_{x}$ in terms
of the closed system, we get more than 500 determining equations for
the functions $X,Y,T,U,\Psi,\Phi,\Psi_{1},\Phi_{1}$ and $P$.
Calculated by computer algebra, the general solutions of them take
the form
\begin{eqnarray}\label{33}
X&=&\frac{1}{3}f_{t}x-\frac{1}{18}f_{tt}y^{2}-
\frac{1}{6}h_{t}y-g,\ \ \ Y=\frac{2}{3}f_{t}y+h,\ \ T=f,
\nonumber\\
U&=&-\frac{2}{3}f_{t}u-\frac{1}{18}f_{tt}x+\frac{1}{108}f_{ttt}y^{2}
+\frac{1}{36}h_{tt}y+\frac{1}{6}g_{t}-c\psi_{1}\phi-c\phi_{1}\psi,\
\nonumber\\
\Psi&=&\left(\frac{1}{324}f_{ttt}y^{3}-\frac{1}{18}xf_{tt}y-\frac{1}{12}xh_{t}
+\frac{1}{72}h_{tt}y^{2}+\frac{1}{6}g_{t}y+F-\frac{cp}{2}\right)\psi,\
\nonumber\\
\Phi&=&\left(\frac{1}{18}xf_{tt}y+\frac{1}{12}xh_{t}-\frac{1}{324}f_{ttt}y^{3}
-\frac{1}{72}h_{tt}y^{2}-\frac{1}{6}g_{t}y-F-\frac13f_t+c_2
-\frac{cp}{2}\right)\phi,\
\nonumber\\
\Psi_{1}&=&\left(
\frac{1}{324}f_{ttt}y^{3}-\frac{1}{12}xh_{t}-\frac{1}{18}xf_{tt}y
+\frac{1}{72}h_{tt}y^{2}+\frac{1}{6}g_{t}y-\frac{1}{3}f_{t}
+F-\frac{cp}{2}\right)\psi_{1}
\nonumber\\
&&-\left(\frac{1}{18}f_{tt}y+\frac{1}{12}h_{t}+\frac{c\phi\psi}{2}\right)\psi,
\nonumber\\
\Phi_{1}&=&\left(
-\frac{1}{324}f_{ttt}y^{3}+\frac{1}{12}xh_{t}+\frac{1}{18}xf_{tt}y
-\frac{1}{72}h_{tt}y^{2}-\frac{1}{6}g_{t}y-\frac{2}{3}f_{t}-F+c2
-\frac{cp}{2}\right)\phi_{1}
\nonumber\\
&&+\frac12\left(\frac{1}{9}f_{tt}y+\frac{1}{6}h_{t}-{c\phi\psi}\right)\phi,
\nonumber\\
P&=&(\frac{1}{3}f_{t}+c_2)p+c_1-\frac{cp^{2}}{2},
\end{eqnarray}
where $\{f,h,g,F\}$ are arbitrary functions of $t$ and
 $\{c,\ c_1,\ c_2\}$ are arbitrary constants. Consequently, it is
convenient to rewrite symmetries (\ref{31}) as
\begin{eqnarray}\label{c011}
\sigma_{u}&=&\left(\frac{1}{3}f_{t}x-\frac{1}{18}f_{tt}y^{2}-
\frac{1}{6}h_{t}y-g\right)u_{x}+\left(\frac{2}{3}f_{t}y+h\right)u_{y}+fu_{t}
+\frac{2}{3}f_{t}u+\frac{1}{18}f_{tt}x-\frac{1}{108}f_{ttt}y^{2}
\nonumber\\
&&
-\frac{1}{36}h_{tt}y-\frac{1}{6}g_{t}
+c(\phi\psi_{1}+\psi\phi_{1}),\
\nonumber\\
\sigma_{\psi}&=&\left(\frac{1}{3}f_{t}x-\frac{1}{18}f_{tt}y^{2}-
\frac{1}{6}h_{t}y-g\right)\psi_{x}+\left(\frac{2}{3}f_{t}y+h\right)\psi_{y}
+f\psi_{t}
+\left(\frac{1}{18}xf_{tt}y+\frac{1}{12}xh_{t}-\frac{1}{324}f_{ttt}y^{3}\right.
\nonumber\\
&&\left.-\frac{1}{72}h_{tt}y^{2}-\frac{1}{6}g_{t}y-F\right)\psi
+\frac12cp\psi,
\nonumber\\
\sigma_{\phi}&=&\left(\frac{1}{3}f_{t}x-\frac{1}{18}f_{tt}y^{2}-
\frac{1}{6}h_{t}y-g\right)\phi_{x}+\left(\frac{2}{3}f_{t}y+h\right)\phi_{y}
+f\phi_{t}
-\left(\frac{1}{18}xf_{tt}y+\frac{1}{12}xh_{t}-\frac{1}{324}f_{ttt}y^{3}\right.
\nonumber\\
&&\left.-\frac{1}{72}h_{tt}y^{2}-\frac{1}{6}g_{t}y-F-\frac13f_t+c_2-\frac12cp\right)\phi,
\nonumber\\
\sigma_{\psi_{1}}&=&\left(\frac{1}{3}f_{t}x-\frac{1}{18}f_{tt}y^{2}-
\frac{1}{6}h_{t}y-g\right)\psi_{1x}+\left(\frac{2}{3}f_{t}y+h\right)\psi_{1y}+f\psi_{1t}
+\left(\frac{1}{18}f_{tt}y+\frac{1}{12}h_{t}\right)\psi
\nonumber\\
&& -\left(
\frac{1}{324}f_{ttt}y^{3}-\frac{1}{12}xh_{t}-\frac{1}{18}xf_{tt}y
+\frac{1}{72}h_{tt}y^{2}+\frac{1}{6}g_{t}y-\frac{1}{3}f_{t}
+F\right)\psi_{1}+\frac12cp\psi_{1}+\frac12c\phi\psi^{2},
\nonumber\\
\sigma_{\phi_{1}}&=&\left(\frac{1}{3}f_{t}x-\frac{1}{18}f_{tt}y^{2}-
\frac{1}{6}h_{t}y-g\right)\phi_{1x}+\left(\frac{2}{3}f_{t}y+h\right)\phi_{1y}
+f\phi_{1t}-
\left(\frac{1}{18}f_{tt}y+\frac{1}{12}h_{t}\right)\phi
\nonumber\\
&&
+\left(\frac{1}{324}f_{ttt}y^{3}-\frac{1}{12}xh_{t}-\frac{1}{18}xf_{tt}y
+\frac{1}{72}h_{tt}y^{2}+\frac{1}{6}g_{t}y+\frac{2}{3}f_{t}+F-c_2
\right)\phi_{1}+\frac12cp\phi_{1}+\frac12c\phi^{2}\psi,
\nonumber\\
\sigma_p&=&\left(\frac{1}{3}f_{t}x-\frac{1}{18}f_{tt}y^{2}-
\frac{1}{6}h_{t}y-g\right)p_{x}+\left(\frac{2}{3}f_{t}y+h\right)p_{y}+fp_{t}
-c_2p-c_1+\frac12cp^{2}.
\end{eqnarray}
On observation of (\ref{c011}), it is not difficult to find that the
$c$-dependent generator of the related Lie symmetry group of Eq.(\ref{21}),
$(\psi_{1}\phi+\phi_{1}\psi)\frac{\partial}{\partial u}$, is just that
corresponding to the Darboux transformation.

Similarity solutions to the closed system can be found by setting
$\sigma_{u},\sigma_{\psi},\sigma_{\phi},\sigma_{\psi_{1}},
\sigma_{\phi_{1}},\sigma_{p}$ be zeros in Eq. (\ref{c011}), which is
equivalent to solving the characteristic equations
\begin{equation}\label{34}
\frac{dx}{X}=\frac{dt}{T}=\frac{du}{U}=\frac{d\psi}{\Psi}
=\frac{d\phi}{\Phi}=\frac{d\psi_{1}}{\Psi_{1}}
=\frac{d\phi_{1}}{\Phi_{1}}=\frac{dp}{P}.
\end{equation}
In $c=0$ case, three reduction cases have been studied by many authors, say, in \cite{KPRed}. In the proceeding paragraphs, three new cases which are generalizations of the known three reductions for $c\neq 0$ are discussed in detail.\\
\textbf{Case 1.} $f\neq0$. In this case, without loss of
generality, we rewrite the arbitrary functions $\{g,\ h,\ F\}$ in
Eq.(\ref{c011}) as
\begin{eqnarray}\label{c19}
h&=&y_{0t}f^{\frac{5}{3}},\qquad f_2\equiv \int f^{-1}\mbox{dt},\nonumber\\  g&=&\frac16f^{7/3}y_{0t}^2-f^{4/3}x_{0t},\nonumber\\
F&=&\frac{1}{48}y_{0t}^3f^3+f_{1t}f+\frac14y_0f(x_{0t}f)_t
+\frac12c_2
\end{eqnarray}
for the expression simplicity with
$x_{0},\ y_{0}$ and $f_{1}$ being arbitrary functions of $t$.

Let us proceed to write the reduction solution $p$ to
$\sigma_p=0$ in the form
\begin{equation}\label{c111}
p=\frac{1}{c}\left\{c_2+2 \Delta \tanh\left[\Delta
(f_{2}+P)\right]\right\},\qquad \Delta=\frac12\sqrt{c_2^2+2cc_1}
\end{equation}
with the group invariant function $P\equiv P(\xi,\ \eta)$ and the similarity variables $\xi,\ \eta$ being given by
\begin{equation}\label{c112}
\xi=x{f^{-\frac{1}{3}}} +\frac{1}{18}y^2f_{t}{f^{-\frac{4}{3}}}
+\frac{1}{6}yy_{0t}f^{\frac{1}{3}}-\frac{1}{18}y_{0}^{2}f_{t}-x_{0},\
\ \ \eta=y{f^{-\frac{2}{3}}}-y_{0}.
\end{equation}
With the help of \eqref{c112}, the group invariant solution for other fields $\psi,\ \phi,\ \psi_1,\ \phi_2$ and $u$ have the forms
\begin{eqnarray}
\psi&=&\Psi\ \mbox{sech}\left[\Delta
(f_{2}+P)\right] e^{-\Omega},\\
\phi&=&\Phi\ \mbox{sech}\left[\Delta
(f_{2}+P)\right]f^{-1/3} e^{\Omega},\\
\psi_1&=&\left\{-\frac{c\Phi\Psi^2}{2\Delta f^{1/3}}
\tanh\left[\Delta(f_{2}+P)\right]-\frac{f_ty\Psi}{18f}
-\frac{f^{2/3}}{12}y_{0t}\Psi+\frac{\Phi_1}{f^{1/3}}\right\}
\mbox{sech}\left[\Delta
(f_{2}+P)\right]e^{-\Omega},\\
\phi_1&=&\left\{-\frac{c\Phi^2\Psi}{2\Delta f^{2/3}}
\tanh\left[\Delta(f_{2}+P)\right]+\frac{f_ty\Phi}{18f}
+\frac{f^{1/3}}{12}y_{0t}\Phi+\frac{\Psi_1}{f^{2/3}}\right\}
\mbox{sech}\left[\Delta
(f_{2}+P)\right]e^{\Omega},\\
u&=&-\frac{f_tx}{18f}+\frac{3ff_{tt}-2f_t^2}{324f^2}y^2
+\frac{(fy_{0t})_t}{36f^{1/3}}y+\frac1{72}y_{0t}^2f^{4/3}-\frac16x_{0t}f^{1/3}\nonumber\\
&&
+\frac{1}{2\Delta^2f^{2/3}}\left\{2U\Delta^2-2c(\Psi\Psi_1+\Phi\Phi_1)
\Delta \tanh\left[\Delta
(f_{2}+P)\right]+c^2\Psi^2\Phi^2\tanh^2\left[\Delta
(f_{2}+P)\right] \right\},\label{c115}
\end{eqnarray}
where $U \equiv U(\xi,\ \eta),\ \Psi\equiv\Psi(\xi,\ \eta), \ \Phi\equiv\Phi(\xi,\ \eta),\ \Psi_1\equiv\Psi_1(\xi,\ \eta)$ and $  \Phi_1\equiv\Phi_1(\xi,\ \eta)$ are further five group invariant functions while $\Omega$ is related to $\{x,\ y,\ t\}$ via
\begin{equation}\label{c122}
\Omega=\frac{2f_ty+3f^{5/3}y_{0t}}{36f}x-\frac{y^3}{324}(\ln f)_{tt}-\frac{y^2}{72}\big(y_{0t}f^{2/3}\big)_t
-\frac{fy_{0t}^2-24x_{0t}}{144}yf^{1/3}-f_1-\frac{y_0x_{0t}}4.
\end{equation}
Now it is straightforward to prove the following first reduction theorem.\\
\bf Theorem 2. \rm If $P$ is a solution of the following reduction equation,
\begin{eqnarray}\label{BP}
&&P_{\xi\xi\xi\xi}P_\xi^2-3P_{\xi\xi\xi}P_\eta^2-4\Delta^2P_{\xi\xi}P_\xi^4
+3P_{\eta\eta}P_\xi^2-(1+4P_{\xi\xi\xi})P_{\xi\xi}P_\xi+3P_{\xi\xi}^2=0,
\end{eqnarray}
then $u$ given by
 \begin{eqnarray}
u&=&-\frac{8\Delta^2P_\xi^4-3P_\eta^2-P_\xi-4P_\xi P_{\xi\xi\xi}+3P_{\xi\xi}^2}{P_\xi^2f^{2/3}}
-\frac{f_tx}{18f}+\frac{3ff_{tt}-2f_t^2}{324f^2}y^2
+\frac{(fy_{0t})_t}{36f^{1/3}}y\nonumber\\
&&
+\frac1{72}y_{0t}^2f^{4/3}
-\frac16x_{0t}f^{1/3}-\frac{2\Delta}{f^{2/3}}\left\{
P_{\xi\xi} -\Delta P_\xi^2\tanh\left[\Delta
(f_{2}+P)\right] \right\}\tanh\left[\Delta
(f_{2}+P)\right],\label{c115a}
\end{eqnarray}
is a solution of the KP equation with $\xi$ and $\eta$ being given by \eqref{c112}.\\
\bf Proof. \rm
Substituting \eqref{c115} into the KP equation \eqref{21} and vanishing the coefficients of the different powers of $\tanh[\Delta(f_2+P)]\equiv T$, one can obtain seven determining equations on six group invariant functions $U,\ P,\ \Psi,\ \Phi,\ \Psi_1$ and $\Phi_1$.
The first determining equation reads
\begin{eqnarray*}
(2\Delta^2P_\xi-c\Psi\Phi)(2\Delta^2P_\xi+c\Psi\Phi)=0,
\end{eqnarray*}
i.e.,
\begin{eqnarray} \label{Psi}
\pm 2\Delta^2P_\xi-c\Psi\Phi=0,
\end{eqnarray}
which comes from the coefficients of $T^6$.
By means of \eqref{Psi}, the second determining equation (the coefficient of $T^5$) is simplified to
\begin{eqnarray}\label{Phi}
c\Phi^2\Phi_1+2\Delta^2\big(\pm\Psi_1P_\xi-\Phi P_{\xi\xi}\big)=0.
\end{eqnarray}
In terms of \eqref{Psi} and \eqref{Phi}, the third determining equation becomes (by vanishing the coefficient of $T^4$)
\begin{eqnarray}\label{U}
U=\frac12(P_{\xi\xi}^2-P_\eta^2)P_\xi^{-2}-\frac43\Delta^2P_\xi^2
+\frac16(1+4P_{\xi\xi\xi})P_\xi^{-1}.
\end{eqnarray}
The fourth determining equation is just the equation \eqref{BP} which is a variant form of the Boussinesq equation and can be read off from the coefficient of $T^3$ after the consideration of \eqref{Psi}, \eqref{Phi} and \eqref{U}. It is also natural that \eqref{BP} is just the reduction equation
of the Schwarzian KP equation \eqref{skp} via \eqref{c111}.

Because the KP equation is Painlev\'e integrable with three resonant conditions being identically satisfied \cite{WTC}, after using the relations \eqref{Psi}, \eqref{Phi} and \eqref{U}, the remained three determining equations given by vanishing the coefficients of $T^2,\ T^1$ and $T^0$ are all identities which have the forms
\begin{equation}\label{c117}
L_{2,1,0}(P,\ \alpha,\ P_\xi,\ P_\eta,\ \cdots,\ \partial_\xi,\ \partial_\eta)\alpha \equiv 0
\end{equation}
where $L_{i},\ i=2,\ 1,\ 0$ are linear operator functions of the indicated variables and
\begin{equation}\label{c117}
\alpha \equiv \mbox{ left \ hand \ side \ of \ \eqref{BP}}.
\end{equation}
Finally, the group invariant solution \eqref{c115} becomes \eqref{c115a} thanks to the relations \eqref{Psi}, \eqref{Phi} and \eqref{U}. The theorem 2 is proved.

From the expression, \eqref{c115} (or \eqref{c115a}), we know that when the DT related part is removed, i.e., $\Delta=0$, the reduction is just the known Boussinesq reduction \cite{KPRed}
\begin{equation}
\big(U_{\xi\xi\xi}-6UU_\xi\big)_\xi+3U_{\eta\eta}=0
\end{equation}
which is related to \eqref{BP} by \eqref{U} for $\Delta=0$. The entrance of the $\tanh$ part in \eqref{c115} (i.e., \eqref{c115a}) means the intrusion of an additional soliton to the Boussinesq wave. In other words, the group invariant solution \eqref{c115a} is an interaction solution of one soliton and one general Boussinesq wave.

\bf Case 2. $f=0,\ h\neq 0$. \rm Because the similar procedure as in Case 1, we can find the second group invariant solution
\begin{eqnarray}\label{c118}
p&=&c^{-1}\left\{c_2+2h\gamma\tanh[\gamma(y+p')]\right\},\quad \gamma=\frac{\sqrt{c_2^2+2cc_1}}{2h},\\
\psi&=&\psi' \mbox{sech}[\gamma(y+p')] e^{\omega},\quad
\omega=\frac{(\ln h)_{tt}}{216h^2}y^3-\frac{h_t}{12h}xy-\frac{ gh_t-2hg_t}{24h^2}y^2+\frac{2F-c_2}{2h}y,\\
\phi&=&\phi' \mbox{sech}[\gamma(y+p')] e^{-\omega},\\
\psi_1&=&\left\{\psi_1'-\frac{\psi'}{12\gamma h}\big[h_t(\gamma y+p')+6c\psi'\phi'\tanh[\gamma(y+p')]\big]\right\} \mbox{sech}[\gamma(y+p')] e^{\omega},\\
\phi_1&=&\left\{\phi_1'+\frac{\phi'}{12\gamma h}\left[h_t(\gamma y+p')+6c\psi'\phi'\tanh[\gamma(y+p')]\right]\right\} \mbox{sech}[\gamma(y+p')] e^{-\omega},\\
u&=&u'-\frac{y+p'}{72h}\left(h_{tt}p'-12g_t-h_{tt}y\right)
-\frac{c}{h\gamma}\big(\psi'\phi_1'+\psi_1'\phi'\big)\tanh[\gamma(y+p')]
\nonumber\\
&&+\frac{c^2}{2h^2\gamma^2}\psi'^2\phi'^2\tanh^2[\gamma(y+p')]
\end{eqnarray}
from $\sigma_u=\sigma_p=\sigma_{\psi}=\sigma_{\phi}=\sigma_{\psi_1}=\sigma_{\phi_1}=0$ with $f=0$, where $u'\equiv u'(x',\ t),\  p'\equiv p'(x',\ t),\  \psi'\equiv \psi'(x',\ t),\ \phi'\equiv \phi'(x',\ t),\ \psi_1'\equiv \psi_1'(x',\ t),\  \phi_1'\equiv \phi_1'(x',\ t)$ and
\begin{equation}\label{c119}
x'\equiv x+\frac{h_ty^2}{12h}+\frac{gy}h.
\end{equation}
Now, it is straightforward to prove the second reduction theorem of the KP equation.\\
\bf Theorem 3. \rm If $p'$ is a solution of the reduction equation
\begin{eqnarray}\label{c123}
&&2(p'_{x'x'x'x'}+p'_{x't})p'^2_{x'}-2p'_tp'_{x'}p'_{x'x'}
-8\gamma^2p'_{x'x'}p'^4_{x'}-(\ln h)_{t}p'^3_{x'}\nonumber\\
&&\qquad +2\left((\ln h)_tp'-4p'_{x'x'x'}-6gh^{-1}\right)p'_{x'x'}p'_{x'}+6p'^3_{x'x'}-6p'_{x'x'}=0,
\end{eqnarray}
then
\begin{eqnarray}\label{c124}
u&=&\frac{p'_t}{6p'_{x'}}-\frac43\gamma^2p'^2_{x'}-\frac{g_tp'}{6h}
+\frac{h_{tt}p'^2}{72h}+\frac{g^2}{2h^2}-\frac{h_tp'-4hp'_{x'x'x'}-g}
{6hp'_{x'}}+\frac{1-p'^2_{x'x'}}{2p'^2_{x'}}\nonumber\\
&& -\frac{y+p'}{72h}(p'h_{tt}-12g_t-yh_{tt})-2\gamma p'_{x'x'}\tanh[\gamma (y+p')]+2\gamma^2p'^2_{x'}\tanh^2[\gamma (y+p')]
\end{eqnarray}
with \eqref{c119} is a solution of the KP equation.

The proof of the theorem 3 will be omitted here because it is similar to the proof of the theorem 2.

From the expression, \eqref{c115} (or \eqref{c115a}), we know that when the DT related part is removed ($c=0$), the reduction will be reduced back the known KdV type reduction \cite{KPRed}
\begin{equation}\label{KdV}
\big(U_{1x'x'x'}-6U_1U_{1x'}\big)_{x'}+U_{1x't}
+\frac{h_t}{2h}U_{1x'}+\frac{h_{tt}}{12h}=0
\end{equation}
with
\begin{equation}\label{KdV}
U_{1}\equiv U-\frac{h_{tt}P^2}{72h}+\frac{g_tP}{6h}-\frac{g^2}{2h^2}.
\end{equation}
The appearance of the $\tanh$ part in \eqref{c124} implies that the group invariant solution is an interaction solution of one soliton and one general KdV type wave.\\
\bf Case 3. $f=h=0$. \rm In this case, the group invariant conditions $\sigma_u=\sigma_p=\sigma_\psi=\sigma_\phi=\sigma_{\psi_1}=\sigma_{\phi_1}=0$ with $f=h=0$ possess the solution
\begin{eqnarray}\label{c126}
p&=&\frac1c\left\{c_2-2g\beta \tanh[\beta(x+q)]\right\},\qquad \beta\equiv
\frac{\sqrt{c_2^2+2cc_1}}{2g},\\
\psi&=& r\ \mbox{sech}[\beta(x+q)] e^{\omega_1},\ \qquad \omega_1\equiv
-\frac{x}{6g}(yg_t+6F-3c_2),\\
\phi&=& s\ \mbox{sech}[\beta(x+q)] e^{-\omega_1},\\
\psi_1&=& \left(\frac{cr^2s}{2g\beta}\tanh[\beta(x+q)]+r_1\right) \mbox{sech}[\beta(x+q)] e^{\omega_1},\\
\phi_1&=& \left(\frac{crs^2}{2g\beta}\tanh[\beta(x+q)]+s_1\right) \mbox{sech}[\beta(x+q)] e^{-\omega_1},\\
u&=&u_1 + \frac{c}{g\beta}(sr_1+rs_1)\tanh[\beta(x+q)]+\frac{c^2s^2r^2}{2g^2\beta^2}
\tanh^2[\beta(x+q)]-\frac{g_t}{6g}(x+q) \label{u1}
\end{eqnarray}
with the group invariant functions $q\equiv q(y,\ t),\ r\equiv r(y,\ t),\ s\equiv s(y,\ t),\ r_1\equiv r_1(y,\ t),\ s_1\equiv s_1(y,\ t)$ and $ u_1\equiv u_1(y,\ t)$.

Similar to the last two cases, substituting \eqref{u1} into the KP equation leads to the following third reduction theorem. \\
\bf Theorem 4. \rm If $q$ is a solution of the reduction equation
\begin{equation}\label{c133}
q_{yy}=\frac{g_t}{3g},
\end{equation}
then
\begin{equation}\label{c132}
u=\frac16q_t+\frac12q_y^2-\frac{4}3\beta^2-\frac{g_t}{6g}(x+q)
+2\beta^2\tanh^2[\beta(x+q)]
\end{equation}
is a solution of the KP equation.

It is clear that the reduction solution \eqref{c132} is corresponding to the usual
single soliton solution (for $g=constant$ and $q$ being a linear function of $y$ and $t$) or the single solitary wave with nonhomogeneous background for $g$ being not a constant.

\section{Interactions among different nonlinear waves}
From the reduction theorems 2,\ 3 and 4 of the last section, one can find  some important aspects:
(1). The DT related symmetries can be used to find new symmetry reductions which are the generalizations and will be reduced back to those of without DT transformations.
(2). The essential role of the DT is to add one soliton no matter what the seed solution is.
(3). The interactions between solitons and other nonlinear waves can be analytically found. The theorem 2 and theorem 3 display the interactions among solitary waves and Boussinesq waves and KdV waves respectively while the third reduction is related to the single solitary waves with homogeneous or nonhomogeneous background.

For concreteness, we write down two further explicit solutions of the first two reduction equations.

For the first reduction equation \eqref{BP}, its cnoidal wave solution is given by,
\begin{equation}\label{Rx1}
R_{x_1}^2-\Delta^2R^4-R+C_1R^2-C_2R^3=0,\ \qquad R\equiv P_{x_1},\ P(\xi,\ \eta)\equiv P(\xi+a\eta)\equiv P(x_1),
\end{equation}
with arbitrary constants $a,\ C_1$ and $C_2$. The solution of \eqref{Rx1} can be explicitly expressed by Jacobi elliptic functions, say,
\begin{equation}\label{R1}
R=-\frac{r_1r_3S^2}{r_1-r_3+r_1S^2},\qquad S\equiv \mbox{sn}\left(\frac12\sqrt{(r_1-r_3)r_2}\Delta x_1,m\right),\quad m\equiv \sqrt{\frac{r_1(r_2-r_3)}{r_2(r_1-r_3)}},
\end{equation}
where $S$ is the Jacobi elliptic function with the modulus $m$ and the arbitrary constants $\{C_1,\  C_2,\ \Delta\}$ have been re-expressed by
\begin{equation}\label{C1C2}
C_1=\frac{r_1r_2+r_1r_3+r_2r_3}{r_1r_2r_3},\ C_2=\frac{r_1+r_2+r_3}{r_1r_2r_3},\
\Delta^2=-\frac{1}{r_1r_2r_3}.
\end{equation}
Correspondingly, the field $P$ has the form
\begin{equation}\label{Pxi}
P=\frac{2r_3}{\Delta \sqrt{r_2(r_1-r_3)}}\left[E_\pi\left(S\sqrt{\frac{r_1-r_3}{r_1-r_3+2r_1S^2}},
\frac{r_1}{r_1-r_3},
m\right)
-E_F\left(S\sqrt{\frac{r_1-r_3}{r_1-r_3+2r_1S^2}},
m\right)\right],
\end{equation}
where $E_F$ and $E_\pi$ are the first and third incomplete elliptic integrals.
It is clear that the exact solution \eqref{c115a} denotes the interaction between the Boussinesq-type cnoidal periodic wave and soliton by taking the arbitrary functions $f,\ y_0$ and $x_0$ as $f=1/v_1,\ x_0=v_2t,\ y_0=v_3(t+t_0)$ with arbitrary constants $v_1,\ v_2,\ v_3$ and $t_0$.

For the second reduction equation \eqref{c123} given in theorem 3, we consider only $h=1,\ g=c_0=constant$ case. For the constant $h$ and $g$ case, its cnoidal wave solution is given by,
\begin{equation}\label{ry1}
r_{y_1}^2-4\gamma^2r^4-1-6c_0R-C_1r^2-C_2r^3=0,\ \qquad r\equiv p'_{y_1},\ p'(x',\ t)=r(x'+a t)\equiv r(y_1),
\end{equation}
where $C_1,\ C_2$ and $a$ are arbitrary constants.

The general solution of \eqref{ry1} can be written as
\begin{equation}\label{R1}
r=\frac{(r_1-r_4)r_3S_1^2-r_4(r_1-r_3)}{r_3-r_1+(r_1-r_4)S_1^2},\ S_1\equiv \mbox{sn}\left(\frac{\sqrt{(r_1-r_3)(r_2-r_4)}}{2r_1r_2r_3r_4} y_1,m\right),\ m\equiv \sqrt{\frac{(r_1-r_4)(r_2-r_3)}{(r_2-r_4)(r_1-r_3)}}
\end{equation}
if the arbitrary constants $\gamma,\ c_0,\ C_1$ and $C_2$ is re-written as
\begin{eqnarray}\label{C1C2c0}
&&C_1=\frac{r_1+r_2+r_3+r_4}{r_1r_2r_3r_4},\
C_2=\frac{r_1(r_2+r_3+r_4)+r_2(r_3+r_4)+r_3r_4}{r_1r_2r_3r_4},\
\nonumber\\
&&\gamma^2=\frac{1}{4r_1r_2r_3r_4},\ c_0=-\frac16\left(\frac1{r_1}+\frac1{r_2}+\frac1{r_3}+\frac1{r_4}\right).
\end{eqnarray}
Consequently, the field $p'$ has the form
\begin{equation}\label{p'}
p'=\sqrt{\frac{4r_1r_2r_3r_4}{ (r_2-r_4)(r_1-r_3)}}\left[(r_3-r_4)E_\pi\left(S_1,
\frac{r_1-r_4}{r_1-r_3},
m\right)
-r_3E_F\left(S,
m\right)\right].
\end{equation}

\section{Discussion and Summary}
To sum up, the non-local symmetry of KP
equation related to the DT can be localized if five potentials, the spectral function $\psi$, the adjoint spectral function $\psi$, the $x$-derivatives of the spectral functions $\{\psi_1\equiv \psi_x,\ \psi_1\equiv \psi_x,\}$ and the singularity manifold function $p\equiv \int \psi\phi \mbox{dx}$, are introduced. For the enlarged system $\{u,\ \psi,\ \phi,\ \psi_1,\ \phi_1,\ p\}$ the generalized Lie point symmetries includes the DT as its special ones.
Using the Lie's first theorem to the Lie point symmetries of the enlarged system, one can find that the first type of DT and the second kind of DT possess the same infinitesimal form but with different group parameters.
From the infinitesimal form of the DT, is is found that the DT of the KP equation is nothing but the M\"obious transformation of its Schwarzian form.
This conclusion is true for the KdV equation \cite{Nonlocal} and may also correct for other integrable systems \cite{Lou}.

Usually, to find the new solutions via DT, one has to solve the spectral problem first. However, for non-constant and non-soliton potential (seed solution) it is very difficult to solve the spectral problem. Fortunately,
In this paper, it is found that it is not necessary to solve the spectral problem directly. The simple alternative way
is to find the symmetry reduction method with the DT related symmetries.
The result shows that if we take the Boussinesq type and the KdV type waves (the general solutions of the Boussinesq and KdV reductions of the KP equation) as seed solutions, the essential and unique role of the DT is to offer an additional soliton.

The KP equation was widely used in various physics fields such as the surface and internal oceanic waves, nolinear nonlinear optics, plasma physics, ferromagnetics, Bose-Einstein condensation, string theory and so on. Though many kind of solutions of the KP equation have been found by many authors via different methods, however, it is difficult interaction solutions among different kinds of nonlinear waves. Especially, the soliton solutions can be used to described various nonlinear phenomena. For instance, the KP's line soliton solutions can be used to describe the propagating tsunami wave. on the other hand, there are many other oceanic surface waves such as the periodic waves which may be described by the Jacobi elliptic periodic solutions (cnoidal waves) of the KP equation. Then how can we find the interaction solutions among the line solitons and the cnoidal waves? In this paper, two special explicit such kinds of solutions are obtained as shown in \eqref{c115a} with \eqref{Pxi} and \eqref{c124} with \eqref{p'} respectively.

The following topics will be discussed in the future series research works: (1).
The method used here can be developed and/or extended to other kinds of nonlocal symmetries such as those of obtained from B\"acklund transformation \cite{Nonlocal}, the bilinear forms and negative hierarchies \cite{neg,XBHu}, the nonlinearizations \cite{Cao}, self-consistent sources \cite{zeng}
 and the Painlev\'e analysis \cite{WTC}. (2). It is shown here that to find the classical symmetry reductions of the enlarged system is equivalent to get the DTs with the seed solutions being taken as the classical symmetry reductions of the original nonlinear system. The method is valid and the conclusion is correct for other kind of complicated seed solutions such as the nonclassical symmetry reductions. (3). The method should and can be applied to other kind of integrable systems, especially for supersymmetric models and discrete ones, to find interaction solutions among different kinds of nonlinear waves. (4). Other types of well known effective methods can also be extended to study the interactions among different kinds of nonlinear excitations.

The work was sponsored by the National Natural Science Foundation of
China (No. 11175092), Scientific Research Fund of Zhejiang
Provincial Education Department under Grant No. Y201017148,
and K. C. Wong Magna Fund in Ningbo University.

\small{
}
\end{document}